\documentclass[a4paper,12pt,notitlepage,onecolumn,oneside]{article}
\usepackage{graphicx,longtable}
\usepackage{epsfig}
\usepackage{cite}
\usepackage{float}
\usepackage{graphicx}
\usepackage{subfigure}
\usepackage{a4wide}

\begin{document}

\title{The soft and the hard pomerons\\
 in hadron elastic scattering at small $t$}
\author{J.R. Cudell$^{a}$, A. Lengyel$^{b}$ and E. Martynov$^{c}$\\~\\
\leftline{\small $^a$ Physique th\'eorique fondamentale, D\'ep. de  Physique, 
Universit\'e de  Li\`ege, }\\
\leftline{\small ~~All\'ee du 6 Ao\^{u}t 17, b\^{a}t. B5a, 
B-4000 Li\`ege~1, Belgium}\\
\leftline{\small ~~E-mail: JR.Cudell@ulg.ac.be}\\
\leftline{\small $^b$ Inst. of Electron Physics, Universitetska 21, UA-88000
Uzhgorod, Ukraine.  
 }\\
\leftline{\small ~~E-mail: sasha@len.uzhgorod.ua}\\
\leftline{\small $^c$ Bogolyubov Inst.
for Theoretical Physics, UA-03143 Kiev, Ukraine.
}\\
\leftline{\small ~~E-mail: martynov@bitp.kiev.ua}\\ 
}
\maketitle

\begin{abstract}
We consider simple-pole descriptions of soft elastic scattering 
for $pp$, $\bar pp$, $\pi^{\pm}p$ and $K^{\pm}p$. We work at
$t$ and $s$ small enough for rescatterings to be neglected, and
allow for the presence of a hard pomeron. After building
and discussing an exhaustive dataset, we show that simple poles 
provide an excellent description of the data in the region 
$- 0.5$ GeV$^2 \leq t \leq -0.1$ GeV$^2$, 6~GeV$\leq\sqrt{s}\leq$ 63~GeV. 
We show that new form factors have to be used, and get
information on the trajectories of the soft and hard pomerons.
\end{abstract}

\section*{Introduction}
In recent papers \cite{clms}, we have shown that a model which includes
a hard pomeron reproduces very well the total cross sections and
the ratios $\rho$ of the real to imaginary parts of the forward 
scattering amplitude,
while the description obtained from a soft pomeron only
is much less convincing \cite{COMPETE}. We considered the full set of forward
data \cite{t0set} for $pp$, $\bar p p$, $Kp$, $\pi p$, $\gamma p$ and
$\gamma\gamma$, and showed that the description extends down to $\sqrt{s}=5$
GeV. 

However, if one uses a simple pole
for the hard pomeron and a fit to all data for $\sqrt{s}\geq 5$~GeV, 
the coupling of this new
trajectory is almost zero in $pp$ scattering, while it is non negligible
in $Kp$ and $\pi p$. The reason is simple: a hard pole, with an intercept
of about 1.45, needs to be unitarised at high energy. Hence the high-energy
$\bar p p$ data almost decouple any fast-rising pole\footnote{This explains the
very small coupling obtained in \cite{DLsig} and the bound of 
\cite{precompete}.}. To see the hard singularity, one thus needs to limit
the energy range of the fit, and we found that for centre-of-mass energies 
5 GeV$\leq \sqrt{s}\leq$ 100 GeV the data were well described by a sum of four simple poles:
a charge-conjugation-odd ($C=-1$) exchange (corresponding to the $\rho$ and 
$\omega$ exchanges and denoted $R_-$) with intercept 0.47, and three $C=+1$ exchanges,
 with intercepts 0.61 ($f$ and $a_2$ trajectories denoted $R_+$), 1.073 (soft pomeron $S$)
and 1.45 (hard pomeron $H$). 

We then showed that it is possible to extend the fit to high energies, 
provided that one unitarises the hard pomeron. The low-energy description
remains dominated by the pole term, whereas the multiple scatterings 
tame the growth at high energy. However, despite the fact that the hard pomeron 
intercept is very close to what is observed in deeply inelastic 
scattering \cite{DisL}
and in photoproduction \cite{DppL}, it is not entirely sure that it is
present in soft scattering. Indeed, its couplings are small and 
its contribution is less than 10\% for $\sqrt{s}<100$~GeV. Hence it
is important to look for confirmation of its presence in other soft
processes, and the obvious place to start from is elastic scattering.

Although elastic scattering has been studied for a long time, its description
within Regge theory poses several problems:
\begin{itemize}
\item
There is no standard dataset: the data are present in the HEPDATA 
system \cite{HEPDATA}, but they have not been gathered into a common format, some of
the included datasets are not published, and several are superseded. Furthermore,
the treatment of systematic errors is often obscure. This may explain
why many authors neglect the quality of their fits: most existing
models do not reproduce the data in a statistically acceptable manner.
\item
Maybe because of the absence of a standard dataset, most theoretical
works concentrate on $pp$ and $\bar p p$ data, and neglect $\pi p$
and $K p$ elastic scattering. As we showed in \cite{clms}, this may
however be the place to look for a hard pomeron.
\item
On the theoretical side, the situation is also more difficult: whereas at
$t=0$ one had to introduce coefficients in front of the Regge exchanges,
one now has to use form factors. These are a priori unknown. Also, there is no
reference fit with an acceptable $\chi^2$ per degree of freedom ($\chi^2$/d.o.f.).
\item
For the purpose of this paper, 
one has to implement several cutoffs: first of all, the energy has to
be sufficient to use leading exchanges, and small enough to be able to
neglect rescatterings\footnote{or to absorb them in the 
parameters describing the simple-pole exchanges.} (especially when we consider contributions from 
a hard pomeron). Similar cut-offs need to be implemented in the 
off-forward case: first of all, many datasets have inconsistencies in the first few bins, so that $|t|$ needs to be large enough\footnote{Besides, one needs to be away from the Coulomb interference
region.}.
At the same time, one needs
to be far from the dip region, where rescatterings are notoriously 
important. Thus there must also be an upper cutoff in $|t|$. 
\end{itemize}

Our strategy in this paper will be to fix the parameters entering the
description of the data at $t=0$ \cite{clms}, and to compare a model
containing only a soft pomeron with a model where we add a hard pomeron.
After a theoretical summary fixing the conventions, 
we shall recall the parametrisation of forward data in section 2.
In section 3, we will present the dataset which we are using,
discuss the problem of systematic errors, and use a 
general method \cite{overlap} to determine the functions describing the
form factors of the various Regge exchanges. As an output, we shall also
be able to determine the position of the first cone in $t$, i.e. the
region where the rescatterings can be neglected.
In section 4, we shall then produce a fit using only a soft pomeron,
and show that it describes very well the elastic data. 
In section 5, we shall give our results for the
hard pomeron case, and give constraints on its form factors and slope.
 
\section{Theoretical framework}
We shall parametrise all exchanges by simple poles, 
and limit ourselves to a region in $s$ and $t$ where these are dominant. 
The amplitude $A^{ab}(s,t)$ that describes the elastic scattering of 
hadrons $a$ and $b$ 
is normalised so that the total and the differential elastic 
cross sections are given by
\begin{eqnarray}
\label{eq:sigtot}
\sigma_{tot}^{ab}(s)&=&\frac{1}{2q_{ab}\sqrt{s}}\Im mA^{ab}(s,0),\\
\frac{d\sigma_{el}^{ab}(s,t)}{dt}&=&\frac{1}{64\pi
sq_{ab}^{2}}|A^{ab}(s,t)|^{2},
\label{eq:sigel}
\end{eqnarray}
where
$q_{ab}=\sqrt{[(s-m_{a}^{2}-m_{b}^{2})^{2}-4m_{a}^{2}m_{b}^{2}]/4s}$
is the momentum of particles $a$ and $b$ in the centre-of-mass
system. 

Regge theory implies that one can write
$A(s,t)\equiv A(z_{t},t)$ where the
Regge variable, $z_{t}$, is the cosine of the scattering angle in the
crossed channel:
\begin{equation}\label{eq:cos}
z_{t}=\frac{t+2s_{ab}}{\sqrt{(4m_{a}^{2}-t)(4m_{b}^{2}-t)}}
\end{equation}
with $s_{ab}=s-m_{a}^{2}-m_{b}^{2}$.

A simple-pole singularity (reggeon) in the complex $j$ plane
at $j=\alpha(t)$ then leads to a term in the amplitude
given by 
\begin{equation}\label{eq:s-pole}
A_{R}^{ab}(z_{t},t)=16\pi^{2}[2\alpha(t)+1]\frac{\Gamma
(\alpha(t)+1/2)}{\sqrt{\pi}\Gamma(\alpha(t)+1)}\beta^{a}(t)\beta^{b}(t)
\eta(\alpha(t))P_{\alpha(t)}(z_{t}),
\end{equation}
where $\alpha(t)$ is the trajectory of the reggeon, $\beta^{i}(t)$ is
the coupling of the reggeon with particle $i$: $t$-channel
unitarity implies that the couplings factorise, and that the
dependence on the beam $a$ and target $b$ enters through the product
$\beta^{a}(t)\beta^{b}(t)$. 
 The signature factor
$\eta(\alpha(t))$ can be written\footnote{We chose the denominators to obtain
 Eqs.~(\ref{eq:sigtot}, \ref{eq:sigel}) automatically, and absorbed their $t$
dependence in $\beta^i(t)$.}
\begin{equation}\label{eq:modsign}
\eta_{\xi}(\alpha(t))=\left
\{
\begin{array}{ll}
\displaystyle -\frac{\exp(-i\pi \alpha(t)/2)}{\sin(\pi
\alpha(0)/2))} &\quad
({\rm crossing\ even},\ C=+1),\\
\displaystyle -i\frac{\exp(-i\pi \alpha(t)/2)}{\cos(\pi
\alpha(0)/2))} &\quad ({\rm crossing\ odd},\ C=-1).
\end{array}
 \right.
\end{equation}

At high energy $s\gg -t$, $z_{t}$ is large. This allows, taking into account
the asymptotics of the Legendre polynomials and using the variable
\begin{equation}\label{eq:zab}
 \tilde s_{ab}=\frac{t+2s_{ab}}{s_{0}}, \qquad \mbox{with} \quad s_{0}=1\ \mbox{GeV}^{2}
\end{equation}
instead of $z_{t}$,
to re-absorb many of the factors of Eq.(\ref{eq:s-pole}) into the
definition of the couplings\footnote{This is in fact necessary if one considers
$\gamma p$ and $\gamma \gamma$ total cross sections for which $t=0$ and 
$m_{a,b}=0$ in Eq.~(\ref{eq:cos}). We also included a factor ${2^{-\alpha_R(0)}}$ so
that the definition of the couplings coincides with that used in [1].}
so that, for the scattering of $a$ on protons, the simple-pole
contribution to the amplitude becomes
\begin{equation}
A_{R}^{ap}(\tilde
s_{ap},t)=\frac{g^{a}_{R}}{2^{\alpha_R(0)}} 
F^a_R(t)F^{p}_{R}(t)\ \eta_{\xi}({ \alpha_{R}(t)})\ 
\tilde s_{ap}^{\alpha_{R}(t)}.
\label{eq:pole}
\end{equation}
with $F^a_R(0)=1$, $a=$ $p$, $\pi$, $K$.

\subsection{Trajectories}

At high enough energies ($\sqrt{s}\geq 5$ GeV \cite{clms}), the amplitude is
dominated by a few exchanged trajectories. 

For the $C=-1$ part, we shall restrict ourselves
to a region in $t$ where it is
enough to consider meson trajectories: one of the reasons to limit ourselves
to the first cone is that we can forget the odderon contribution, which is known to
be negligible at $t=0$. 

For the $C=+1$ part, we shall first consider meson exchanges, as well
as a soft pomeron and a hard pomeron. 

We shall consider here scattering of $p$, $\bar p$, $\pi^\pm$ and $K^\pm$ on protons,
and we summarise the possible exchanged trajectories in Table~\ref{tab:trajs}.
\begin{table}
\begin{center}
\begin{tabular}[h]{|c|c|c|c|}
\hline
$a$       &  $C=+1$ & $C=-1$ & $A^{ap}(s_{ap},t)$\\\hline
$p$       &  $P, f, a_{2}$ & $ \omega ,\rho $ & $A^{pp}=P+f+a_{2}-\omega-\rho ,$\\
$\bar p$  &                &                     & $A^{\bar p p}=P+f+a_{2}+\omega+\rho ,$\\\hline
$\pi^{+}$ &  $P, f$ & $\rho $ & $A^{\pi^{+}p}=P+f-\rho ,$\\
$\pi^{-}$ &         &          & $A^{\pi^{-}p}=P+f+\rho ,$\\\hline
$K^{+}$   & $P, f, a_{2}$ & $\omega ,\rho $ & $A^{K^{+}p}=P+f+a_{2}-\omega-\rho ,$\\
$K^{-}$   &               &                 & $A^{K^{-}p}=P+f+a_{2}+\omega+\rho ,$\\\hline
\end{tabular}
\end{center}
\caption{The trajectories entering the amplitudes considered in this paper.}
\label{tab:trajs}
\end{table}

Generally, the $\omega $, $\rho$, $f$ and $a_2$ trajectories are different: 
they do not have coinciding intercepts or
slopes\cite{dgmp}. However, as each trajectory comes with three form
factors, we shall have to assume degeneracy for the $C=+1$
and for the $C=-1$ trajectories \cite{CKK}, in order to limit the number of 
parameters. 

Hence the model that we are considering can be written:
\begin{equation}
A^{ap}(s,t)=A_+^{ap}(\tilde s_{ap},t)+A_S^{ap}(\tilde s_{ap},t)+A_H^{ap}(\tilde s_{ap},t)\mp A^{ap}_-(\tilde s_{ap})
\label{eq:amplitude}
\end{equation}
with the $-$ sign for the (positively charged) particles.

\section{Description of the forward data}
We have shown in \cite{clms} that the data for $pp$, $\bar p p$, $\pi^\pm p$, $K^\pm p$,
$\gamma p$ and $\gamma\gamma$ can be well described from $\sqrt{s}=5$ GeV
to\footnote{The hadron-hadron data extend to 62.4 GeV.} 100 GeV
 by either a soft pomeron, or a mixture of a soft pomeron and a hard pomeron,
the latter case being significantly better. We have also shown that the inclusion
of the subtraction constants that enter the dispersion relations lead to
a better description of the real part of the amplitude. The formula for the $\rho$ parameter
is then given by 
\begin{equation}
\label{eq:final dr}
\rho _{\pm }\, \sigma _{\pm }=\frac{R_{ap}}{p}+\frac{E}{\pi p}{\textrm{P}}\int
_{m_{a}}^{\infty }\left[ \frac{\sigma _{\pm }}{E'(E'-E)}-\frac{\sigma _{\mp
}}{E'(E'+E)}\right] p'\, dE'
\end{equation}
where the \( + \) sign refers to the process \( ap\rightarrow ap \)
and the \( - \) sign to \( \bar{a}p\rightarrow \bar{a}p \), \( E \)
and $p$ are the energy and the momentum 
of $a$ in the proton rest frame, P indicates 
a principal-part integral, \( R_{ap} \) is the subtraction constant,
and $\sigma$ are the total cross sections. They are given 
by Eqs.~(\ref{eq:sigtot}, \ref{eq:amplitude}) for $\sqrt{s}\geq 5$ GeV,
and fitted directly to the data at lower energy \cite{clms}.

We give in Table 2 the values of the parameters resulting for a fit to
all data for $\sigma_{tot}$ and $\rho$ for $\bar p p$, $pp$, $\pi^\pm p$ and $K^\pm p$,
and for  $\sigma_{tot}$ for $\gamma p$ and\footnote{We have used
the factorisation of the simple-pole residues to obtain the amplitude for $\gamma\gamma$ \cite{clms}.} $\gamma\gamma$. 
We quote the values obtained in \cite{clms} 
(for a model with both a soft and a hard
pomeron), and follow the same procedure for a model with a soft pomeron only. 
Table~\ref{tab:chi-0} shows the quality of the fits. Clearly, even in this
 modest energy range,
the inclusion of a hard pomeron makes the fits much better, particularly those to 
the $\rho$ parameter for pions and kaons. 
Converting the $\chi^2$/d.o.f.
into a confidence level (CL), one gets for the overall soft pomeron CL=6\%, 
whereas the fit including a hard pomeron achieves CL=93\%. Nevertheless, as the existence of the hard
pomeron is not totally settled, we shall
keep both models in the following, and see how well they fare in the description of the elastic
data.
\begin{table}
\begin{center}
\begin{tabular}{|c|c|c|}
\hline
 Parameter             & soft pomeron & soft \& hard pomerons \\
\hline
       $\alpha_{S}(0)$       & 1.0927 & 1.0728 \\
       $\alpha_{H}(0)$       & -      & 1.45  \\
        $\alpha_{+}(0)$      & 0.61   & 0.61\\
       $\alpha_{-}(0)$  & 0.47   & 0.47   \\\hline
        $g_{H}^{p}$        & -      & 0.10  \\
       $g_{H}^{\pi}$       & -      & 0.28  \\
        $g_{H}^{K}$        & -      & 0.30  \\\hline
        $g_{S}^{p}$        & 49.5  & 56.2 \\
      $g_{S}^{\pi}$       & 31.4  & 32.7 \\
       $g_{S}^{K}$        & 27.7  & 28.3 \\\hline
       $g_{+}^{p}$        & 177  & 158 \\
      $g_{+}^{\pi}$       & 78  & 78 \\
       $g_{+}^{K}$        & 43  & 46 \\\hline
      $g_{-}^{p}$    & 81  & 79 \\
     $g_{-}^{\pi}$   & 13.9  & 14.2 \\
      $g_{-}^{K}$    & 32  & 32 \\
\hline
\end{tabular}
\end{center}
  \caption{Values of the intercepts and couplings ($t=0$).}
\label{tab:teq}
\end{table}

\bigskip

\begin{table}
\begin{center}
\begin{tabular}{|c|c||c||c|}
\hline Quantity     &  Number          & soft & soft and hard \\
                    &  of points  $N$  &$\chi^{2}/N$  & $\chi^{2}/N$ \\
\hline
$\sigma_{tot}^{pp}$             & 104 &  1.2 &  0.86  \\
$\sigma_{tot}^{\bar pp}$        &  59 &  0.78 &  0.88  \\
$\sigma_{tot}^{\pi^{+}p}$       &  50 &  1.2 &  0.78  \\
$\sigma_{tot}^{\pi^{-}p}$       &  95 &  0.90 &  0.90  \\
$\sigma_{tot}^{K^{+}p}$         &  40 &  0.93 &  0.72  \\
$\sigma_{tot}^{K^{-}p}$         &  63 &  0.72 &  0.62  \\
$\sigma_{tot}^{\gamma p}$       &  38 &  0.61 &  0.57  \\
$\sigma_{tot}^{\gamma \gamma}$  &  34 &  0.87 &  0.74  \\
\hline
$\rho^{pp}$             &  64 & 1.59 & 1.62  \\
$\rho^{\bar pp}$        &   9 & 0.49 & 0.43  \\
$\rho^{\pi^{+}p}$       &   8 & 2.8 &  1.52  \\
$\rho^{\pi^{-}p}$       &  30 & 1.8 &  1.09  \\
$\rho^{K^{+}p}$         &  10 & 0.72 &  0.70  \\
$\rho^{K^{-}p}$         &   8 & 1.7 &   0.90  \\
\hline
Total                 & 603 & 1.07 &  0.95 \\
\hline
\end{tabular}
\end{center}
  \caption{Partial $\chi^{2}$ for the total cross sections
  $\sigma_{tot}$ and the ratios $\rho$.}\label{tab:chi-0}
\end{table}

\section{The elastic dataset}
Throughout the last 40 years, there have been many measurements of the
differential elastic cross sections [\citen{ABE}-\citen{SCHIZ}]. In the present paper,
we shall use not only $pp$ and $\bar pp$ data, but also $K^\pm p$ and $\pi^\pm p$
data as the hard pomeron seems to couple more to mesons \cite{clms}. Fortunately,
most of these measurements have been communicated to the HEPDATA group \cite{HEPDATA},
so that one does not need to re-encode all the data. However, some basic work still
needs to be done, as there are 80 papers, with different conventions,
and various units. Once the translation into a common format has been achieved,
there are still a number of issues to be dealt with:
\begin{itemize}
\item Some of the data are preliminary or redundant. We chose to include
only final published data in the set that we are building;
\item The main systematic error usually comes from a poor knowledge of the beam 
luminosity. This means that all the data of one run taken in a given experiment
at a given energy can be shifted up or down by a certain amount. Although we shall mostly treat
these errors as random (and add them quadratically to the statistical error), we
have encoded this information in the dataset. 
Hence we have split the data into
subsets, to which correspond data in a given paper 
with the same systematic error. This defines 263 different subsets of the data, shown
in Appendix~1. We shall also use this information to exclude subsets
which blatantly contradict the rest of the dataset.
\item Several experiments have not spelled out their systematic errors in 
the published work, and these have to be reconstructed. Indeed, many  
measurements are not absolute, but rather normalised by
extrapolating to the optical point $d\sigma_{el}/dt(t=0)$, which is known from 
measurements of the total cross section. In that case, we have assigned 
the error on the optical point ({\i.e.} twice that on the total cross section
used) as a systematic error on the subset. 
\item In the case of bubble chamber experiments, such as \cite{BRICK}, the 
luminosity was monitored, but it was included in the systematic uncertainty added to the
statistical one. We have thus subtracted it so that these data can be shifted
in the same way as the others.
\item In the case of \cite{SCHIZ}, we have added the $t$-dependent systematics
to the statistical error, and allowed 4\% in the global normalisation.
\item As we shall see in the subsequent sections of this paper, 
some of the subsets \cite{BRUNETON,BOGOLYUBSKY,ARMITAGE,AKERLOF} are
 in strong disagreement with the other sets considered. We shall 
eventually exclude them from our analysis.
\end{itemize}
The global dataset \cite{Edata} contains 10188 points (we have restricted it to
data at $\sqrt{s}\geq 4$ GeV). We show some of its
details in the tables of the Appendix. The
present analysis, which concentrates on the first cone,
 will include about a fourth of these data, as explained in the next section,
and shown in Table~\ref{tab:stats}.

\begin{table}
\begin{center}
\begin{tabular}{|c|c|c|c|c|c|c|c|}
\hline
observable&$N_{pp}$&$N_{\bar p p}$&$N_{\pi^+ p}$&$N_{\pi^- p}$&$N_{K^+ p}$&$N_{K^-p}$&$N_{tot}$\\
\hline
 $\sigma_{tot}$  (full set, all $\sqrt{s}$)      & 261& 444 & 412& 606 & 208& 416&2347\\
                   this analysis& 104& 50  &50  &95   &40  &  63&402\\
\hline
 $\rho$            (full set, all $\sqrt{s}$)     &116 & 90  & 9  & 39  & 22 & 15&291\\
                   this analysis&64  &9   &8   &30   &10  & 8&129\\
\hline
 $d\sigma_{el}/dt$ (full set, $\sqrt{s}\geq 4$ GeV)     &4639& 1252& 802& 2169& 595& 731&10188\\
          this analysis&818 &281  &290 &483  &166 &169&2207\\
          after exclusion&795 &226  &281 &478  &166 &169&2115\\
\hline
\end{tabular}
\end{center}
\caption{The statistics of the full dataset and of the present analysis.}
\label{tab:stats}
\end{table}
The forward fit of section 1 gave us the intercepts and the couplings $g_\pm$,
$g_S$ and $g_H$. To extend it to non-zero $t$, we need to find the form
factors. These are 
{\it a priori} unknown, so that one has to deal with arbitrary functions.
\subsection{Form factors and local fits}
In order to obtain the possible form factors, we shall scan the dataset
at fixed $t$, {\it i.e.} we shall fit a complex
amplitude 
with constant form factors to the data in small bins of $t$ 
(and refer to these fits as $local$ fits)\footnote
{Note
that we shall neglect the subtraction constants 
of the real part in the following. We checked that their inclusion does not
significantly improve the description of non-forward data.}.
The constants that we get will then depend on $t$
and give us a picture of the form factor. The value of the $\chi^2$ will also tell us
in which region of $t$ we should work. 

This strategy however will not
work for the general case considered here: each bin does not contain enough points
to have a unique minimum. We can take advantage of the fact that both models considered here
give the same values
for the intercept of the crossing-odd Reggeon contribution,
and for the crossing-even ones as well 
(see Table~\ref{tab:teq}).
We can also read off the slopes from a Chew-Frautschi plot.
This gives the following $f/a_2$ and $\rho/\omega$
trajectories:
\begin{eqnarray}
\alpha_+&=& 0.61+0.82~t,\nonumber\\
\alpha_-&=&  0.47+0.91~t.
\label{eq:Chew}
\end{eqnarray}
Furthermore, we shall not be able to include a hard pomeron in the local fits
as its contribution is too small to be stable.

We fit all the data from 6 GeV$\leq\sqrt{s}\leq 63$ GeV, and we choose small bins 
of width 0.02~GeV$^2$. 
We restrict ourselves to independent 
bins where we have more than four points for each process.
\begin{figure}
\includegraphics[scale=.7]{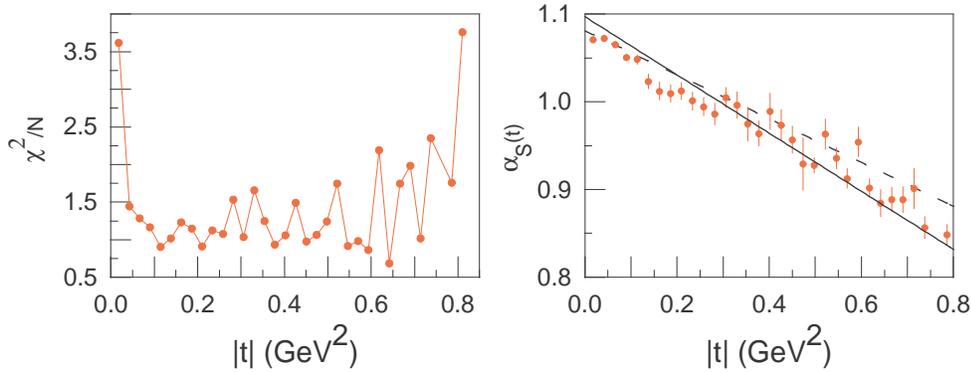}
\caption{The results of the local fits for the $\chi^2$ per number
of points (left) and for the pomeron trajectory (right).
The dashed curve is from \cite{DLel} and the plain curve 
results from the global fit given in the next section.} 
\label{fig:tbins1}
\end{figure}
\begin{figure}
\begin{center}
\includegraphics[scale=.7]{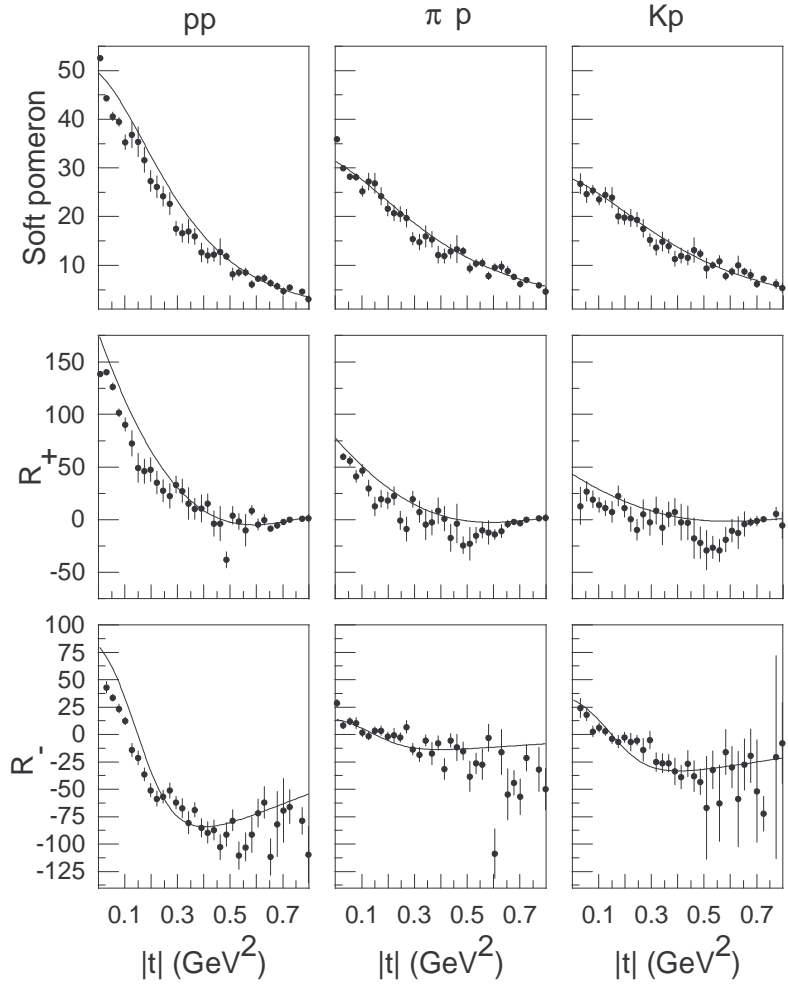}
\end{center}
\caption{The results of the local fits for the residues of the poles.
The curves are the results of a global fit explained 
in the next section.} 
\label{fig:tbins2}
\end{figure}

Each of these fits gives us a values of the $\chi^2$ per number of points,
the coefficients $g_R^{ap}F^p_R(t)F^a_R(t)$, as well as
$\alpha_S(t)$ for each $t$. 
We show these results in Figs.~\ref{fig:tbins1} and \ref{fig:tbins2}.
The $\chi^2$ curve of Fig.~\ref{fig:tbins1} shows two things: 
first of all, the fit is never perfect, and this
can be traced back to incompatibilities in the data\footnote{The inclusion of data for $\sqrt{s}\leq 6$ GeV would only make this problem worse.}. 
We shall come back to this in the next section, when we perform a global fit to
all data. 
The second lesson is that the simple-pole description of the data has a chance to
succeed in a limited region: the $\chi^2$ grows fast both at low $|t|$ (partly because
the Coulomb interaction begins to matter)
and for $|t|>0.6$ (where multiple exchanges come into play). 
To be conservative, we shall consider
in the following\footnote{ 
We have tried several possibilities
for the meson trajectories, and also added a hard pomeron to the local fits. The
range of validity of the fit is not affected by these details.}
the region $0.1\leq |t|\leq 0.5$.
The right-hand graph in Fig.~\ref{fig:tbins1} shows the soft pomeron trajectory. It is
very linear as a function of $t$. Its intercept and slope are somewhat different
from the standard ones \cite{DLel}.

Figure \ref{fig:tbins2} shows the results for the residues of the poles
$g^a F_{R}^a(t)F_{R}^p(t)$. 
In all cases, it is obvious that form factors must be different
for different trajectories. There is in fact no reason why the
hadrons should respond in the same way to different exchanges,
as these have different quantum numbers and different ranges, and 
couple differently to quarks and gluons.

For the soft pomeron, we find that  we can get a good description in 
the $pp$ and $\bar p p$ cases
if we take
\begin{equation}\label{eq:sffp2}
F^{p}_{S}(t)=\frac{1}{1-t/t_{S}^{(1)}+\left(t/t_{S}^{(2)}\right)^2}.
\end{equation}
For $\pi$ and $K$ mesons, an adequate fit is provided by
 the monopole form factors\footnote{although in this limited range of $t$ 
it is also possible to use dipoles.}
\begin{equation}
F^{a}_{S}(t)=\frac{1}{1-t/t^{a}_{S}}, \qquad a=\pi, K. 
\label{eq:softc}
\end{equation}

The local fits for both the $C=+1$ and the $C=-1$ reggeons indicate that
the form factors have a zero at some $t$ value. In the crossing-odd case,
this is the well-known cross-over phenomenon \cite{crossovers}: 
the curves for $d\sigma/dt$ for $pa$ and $p\bar a$ cross
each other at some value of $t$. In the crossing-even case, the zero
is close to the upper value of $|t|$, so that we have evidence for a sharp
decrease but not necessarily for a zero.

In each case, we have tried to obtain such zeroes through rescatterings.
However, it is hard then to cancel both the real and the imaginary parts,
and the zero moves with energy, or disappears when energy changes. We thus
assume here, in a way which is consistent with the simple-pole hypothesis,
that these zeroes are the same for $pp$, $\bar p p$, $\pi^\pm p$ and $K^\pm p$
scattering, and that they are fixed with energy: they can be thought of
as a property of the form factors, or of the exchange itself, and are consistent
with Regge factorisation.

We thus parametrise the $R_-$ and $R_+$ contributions as
\begin{equation}
A_{\pm}^{ap}(\tilde
s_{ap},t)=Z^{a}_{\pm}(t)g^{a}_{\pm}F^a_{\pm}(t)F^{p}_{\pm}(t)\ 
\eta_{\xi}({ \alpha_{\pm}(t)})\ 
\tilde s_{ap}^{\alpha_{\pm}(t)}.
\label{eq:pole-}
\end{equation}
For the form factors $F^{a}_{\pm}(t)$, we take the form
\begin{equation}\label{eq:fp}
F^{p}_{\pm}(t)=\frac{1}{\left(1-t/t^p_{\pm}\right)^2}.
\end{equation}
in the proton case, whereas we find that
\begin{equation}\label{eq:fpik}
F^{\pi,K}_{\pm}(t)=F_S^{\pi,K}(t)
\end{equation}
gives us a good fit for $\pi$ and $K$.

The factor
$Z^{a}_{\pm}(t)$  
has a common zero $\zeta_\pm$, independent of $s$, for $p,\pi,K$, 
but a different one for the $C=+1$ and
the $C=-1$ trajectories: 
\begin{equation}\label{eq:oz-fixed}
Z^{a}_{\pm}(t)=
\frac{\tanh(1+t/\zeta_{\pm})}{\tanh(1)}, 
\quad a=p,\pi,K.
\end{equation}
We choose this simple form to restrict the growth of $Z^{a}_{\pm}$ with $t$.

Finally, when we shall introduce a hard pomeron, we shall find that a 
dipole form factor describes the proton data well
\begin{equation}\label{eq:fh}
F^{p}_{H}(t)=\frac{1}{\left(1-t/t^p_{H}\right)^2},
\end{equation}
 whereas we can use the same
form factor as for the soft pomeron to describe pions and kaons:
\begin{equation}\label{eq:fpikh}
F^{\pi,K}_{H}(t)=F_S^{\pi,K}(t).
\end{equation}

We summarise in Table~\ref{tab:forms} our choice of form factors.
Of course, these are the simplest functions that reproduce the data
at the values of $t$ considered here. Consideration of different $t$
ranges will probably call for more complicated parametrisations.
\def\dst{\displaystyle}
\begin{table}
\label{tab:forms}
\begin{center}
\begin{tabular}{c|ccc}
    & $p$ & $\pi$ & $K$ \\
      \hline
 & & &     \\
 $S$    & $\dst \frac{1}{1-t/t_{S}^{(1)}+(t/t_{S}^{(2)})^{2}}$  & $\dst\frac{1}{1-t/t^{\pi}}$ & $\dst\frac{1}{1-t/t^{K}}$ \\
& & & \\
$C=+1$ &$\dst \frac{1}{(1-t/t_{+})^{2}}$   & $\dst\frac{1}{1-t/t^{\pi}}$  & $\dst\frac{1}{1-t/t^{K}}$ \\
 & & &     \\
$C=-1$ &$\dst \frac{1}{(1-t/t_{-})^{2}}$   & $\dst\frac{1}{1-t/t^{\pi}}$  & $\dst\frac{1}{1-t/t^{K}}$ \\
 & & &     \\
$H$ &$\dst \frac{1}{(1-t/t_{H})^{2}}$  & $\dst\frac{1}{1-t/t^{\pi}}$   & $\dst\frac{1}{1-t/t^{K}}$ \\
\end{tabular}
\end{center}
  \caption{Parametrisation of the form factors.}
\end{table}

\section{Soft pomeron fit}
Equipped with the information from the local fits, we can now perform a global
fit to the elastic data for 0.1 GeV$^2\leq |t|\leq$ 0.5 GeV$^2$, for
6 GeV$\leq \sqrt{s}\leq $ 63 GeV, and for a soft pomeron only. 
We fix the trajectories of the $C=+1$ and $C=-1$ exchanges according to 
Eq.~(\ref{eq:Chew}). 

The $\chi^2$/d.o.f. reaches the value 1.45, which is unacceptable
for the number of points fitted (2207). Such a high value of the $\chi^2$ is
largely due to contradictions between sets of data.
\begin{table}
\begin{center}
\begin{tabular}{|c|c|c|}
\hline
Parameter & soft pomeron & soft and hard pomerons \\
\hline
 $\alpha\, '_{S}$  (GeV$^{-2}$)  & 0.332 $\pm$ 0.007  & 0.297 $\pm$ 0.010  \\
 $\alpha\, '_{H}$  (GeV$^{-2}$)  & -                  & 0.10 $\pm$ 0.21   \\
  $\alpha\, '_{+}$ (GeV$^{-2}$)  & 0.82 (fixed)       & 0.82 (fixed)       \\
 $\alpha\, '_{-}$  (GeV$^{-2}$)  & 0.91 (fixed)       & 0.91 (fixed)       \\
   $t_{S}^{(1)}$  (GeV$^{2}$)   & 0.56 $\pm$ 0.01    & 0.56 $\pm$ 0.02    \\
   $t_{S}^{(2)}$  (GeV$^{2}$)   & 2.33 $\pm$ 0.34    & 1.16 $\pm$ 0.06  \\
  $t_{H}$          (GeV$^{2}$)   & -                  & 0.20 $\pm$ 0.05    \\
  $t_{+}$          (GeV$^{2}$)   & 2.96 $\pm$ 0.25    & 2.34 $\pm$ 0.22    \\
   $t_{-}$         (GeV$^{2}$)   & 7.97 $\pm$ 1.41    & 9.0 $\pm$  1.8      \\
  $t^{\pi}$        (GeV$^{2}$)   & 2.53 $\pm$ 0.14    & 2.89 $\pm$ 0.23    \\
  $t^{K}$          (GeV$^{2}$)   & 3.92 $\pm$ 0.28    & 6.33 $\pm$ 0.94    \\
  $\zeta_{-}$      (GeV$^{2}$)   & 0.148 $\pm$ 0.003  & 0.153 $\pm$ 0.003  \\
 $\zeta_{+}$       (GeV$^{2}$)   & 0.47 $\pm$ 0.02    & 0.47 $\pm$ 0.03    \\
\hline
\end{tabular}
\end{center}
  \caption{Values of the parameters (fit at $t\neq 0$).}\label{tab:fits}
\end{table}
We thus excluded the following data, which all have a CL less than $10^{-8}$:
Bruneton \cite{BRUNETON} (sets 1050, 1204 and 1313, 25 points), Armitage 
\cite{ARMITAGE} (set 1038, 12 points), Akerlof \cite{AKERLOF} $\bar p p$ 
for $\sqrt{s}= 9.78$ GeV (set 1101, 20 points) 
and Bogolyubsky \cite{BOGOLYUBSKY} (set 1114, 35 points).
The removal of these 92 points (less than 5\% of the data) 
brings the $\chi^2$/d.o.f. to 1.03, i.e. a confidence level of 20\%.

The parameters of the fit are given in Table~\ref{tab:fits}, and the partial $\chi^2$
in Table~\ref{tab:chis}. We also show the form factors resulting from the global fit in
Fig.~\ref{fig:tbins2}. We see that there is good agreement with the local fits. 

The main result is that
the slope of the soft pomeron is higher than usually believed: $\alpha'_S\approx 0.3$ 
GeV$^{-2}$. 
Also, the fit to near-forward data is remarkably good\footnote{The 
fact that the soft pomeron reproduces elastic scattering well 
while it fails to reproduce data at $t=0$ is due to the very different
systematic errors, which are typically of a few percents in forward 
data, and of order 10\% in elastic near-forward data.}. 

\begin{figure}[H]
\centerline{\includegraphics[scale=0.7]{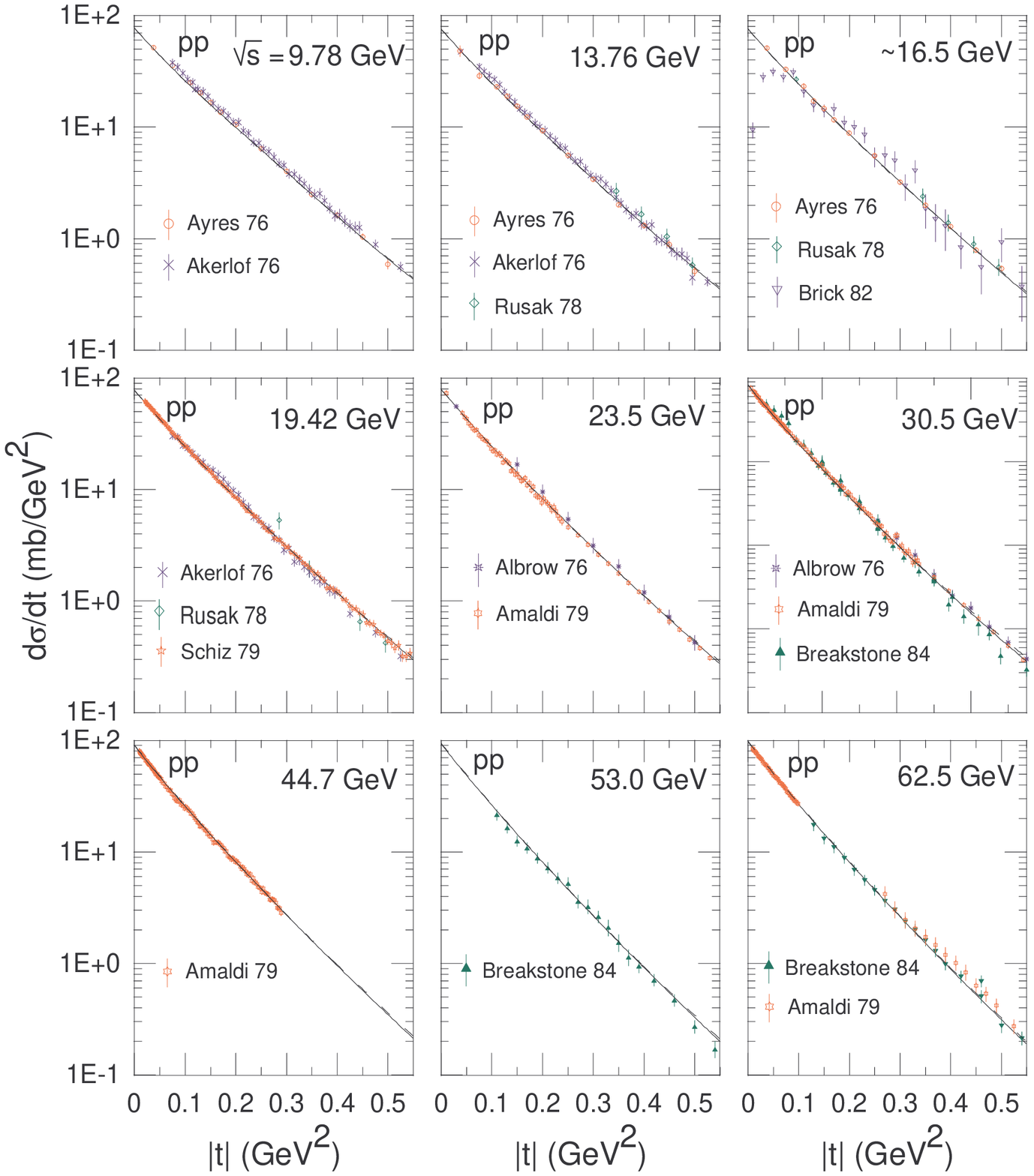}}
\caption{$pp$ differential cross sections. The plain curve shows the soft pomeron fit, and the dashed one the fit that includes a hard pomeron.} \label{fig:pp}
\end{figure}

\begin{figure}[H]
\centerline{\includegraphics[scale=0.7]{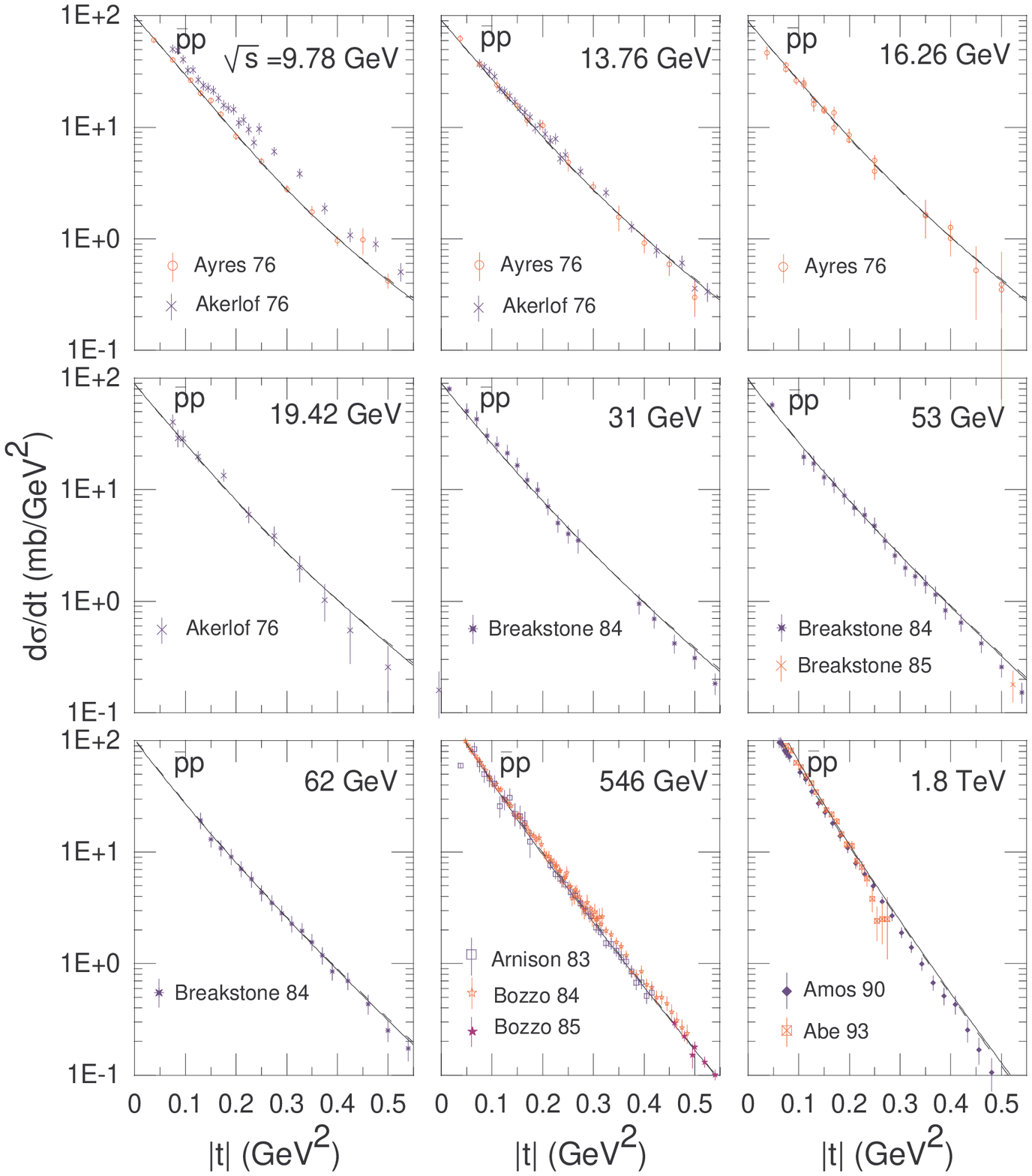}}
\caption{$p\bar p$ differential cross sections. The plain curve shows the soft pomeron fit, and the dashed one the fit that includes a hard pomeron.} \label{fig:pbarp}
\end{figure}

\begin{figure}[H]
\includegraphics[scale=0.50]{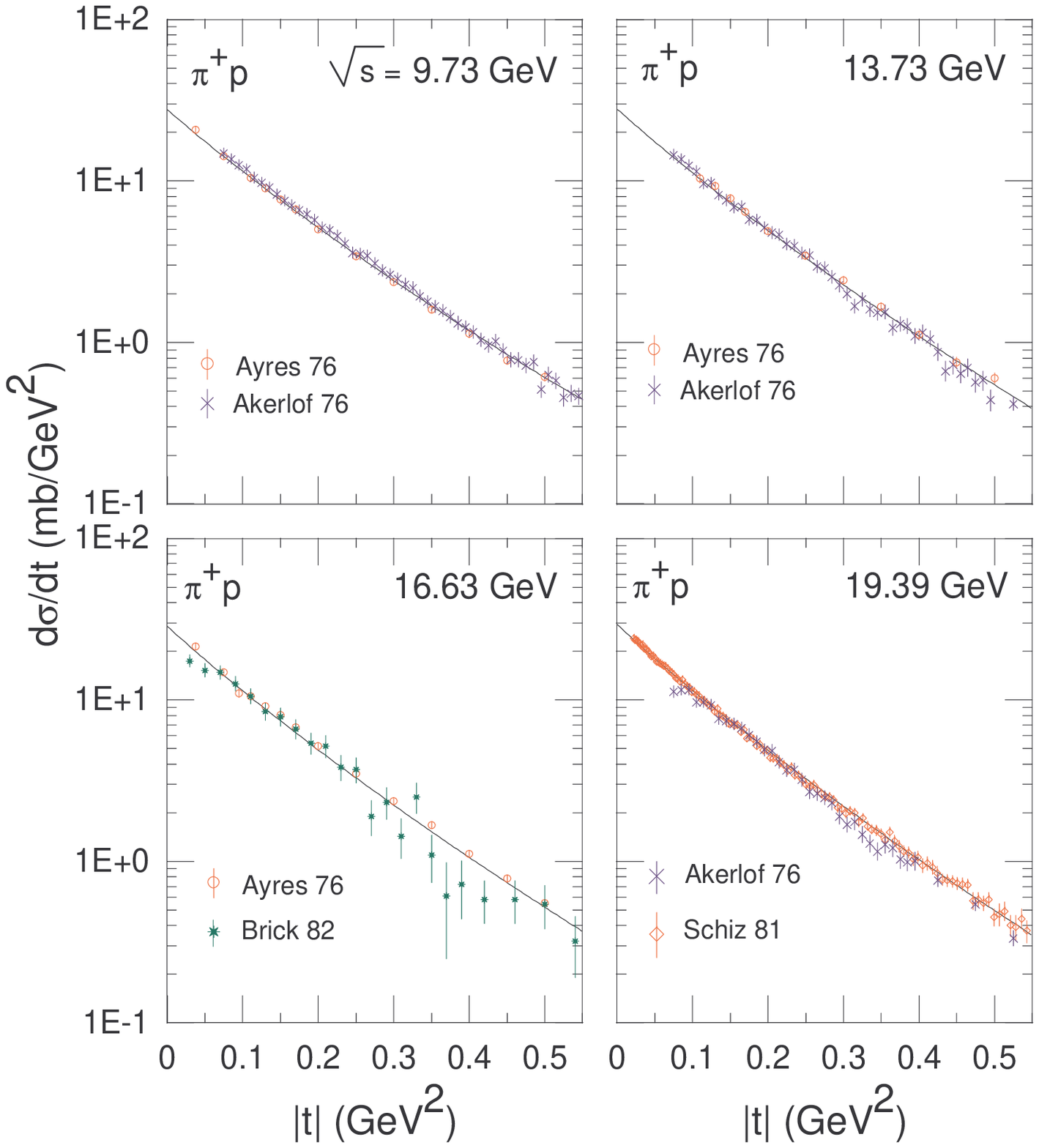}
\includegraphics[scale=0.50]{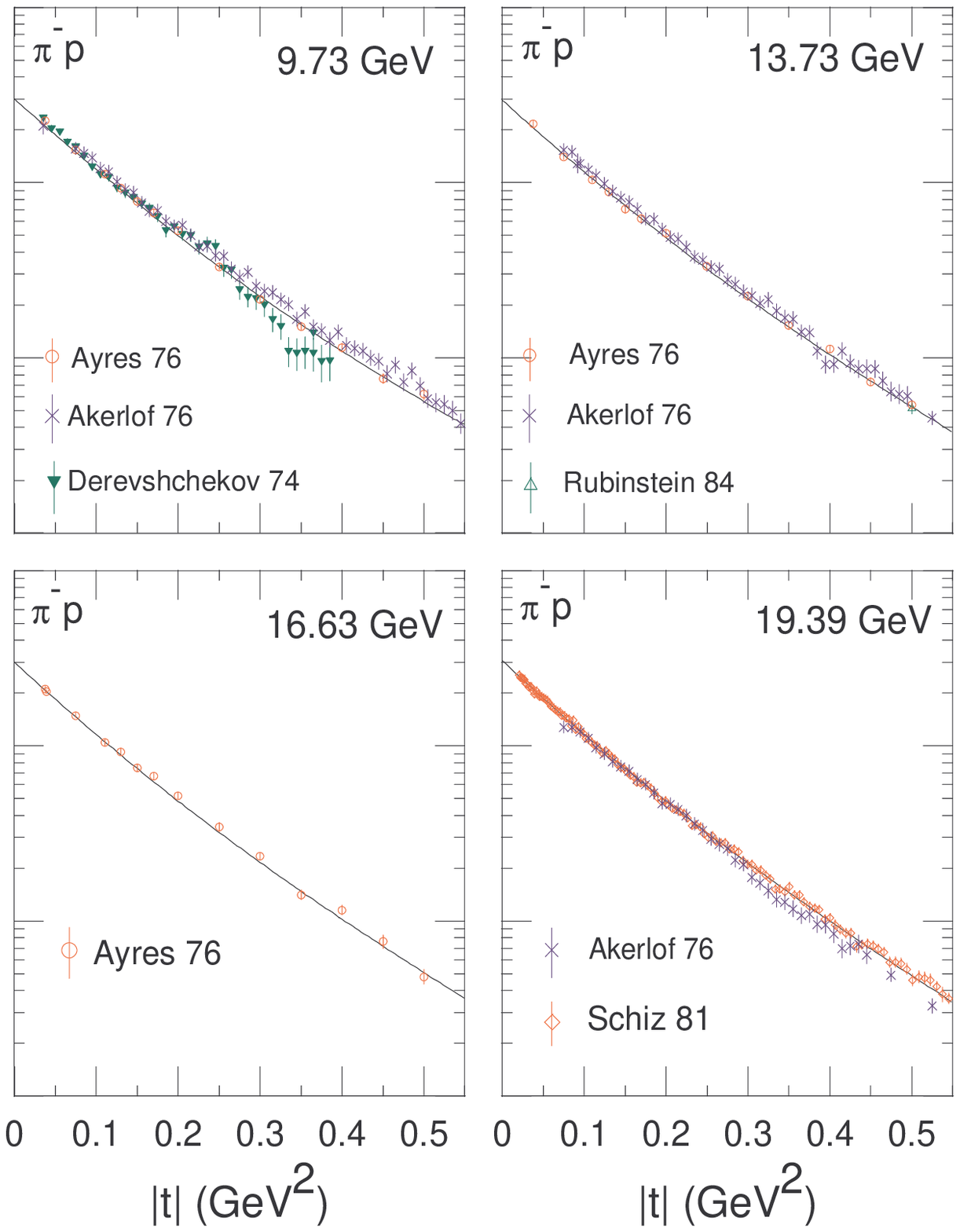}
\caption{$\pi^{+}p$ and $\pi^{-}p$ differential cross sections.}
\label{fig:pi}
\end{figure}

\begin{figure}[H]
\includegraphics[scale=0.50]{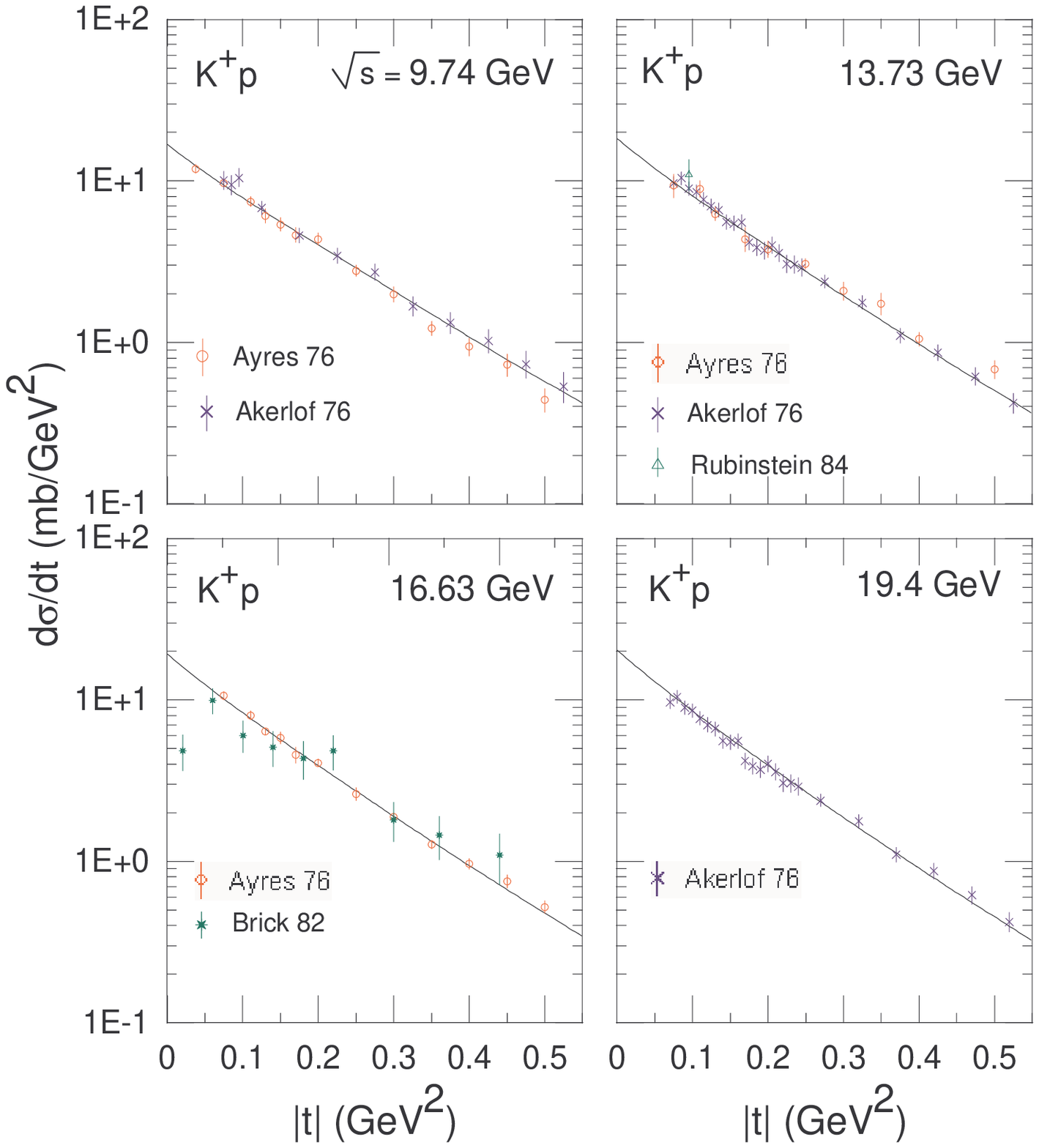}
\includegraphics[scale=0.50]{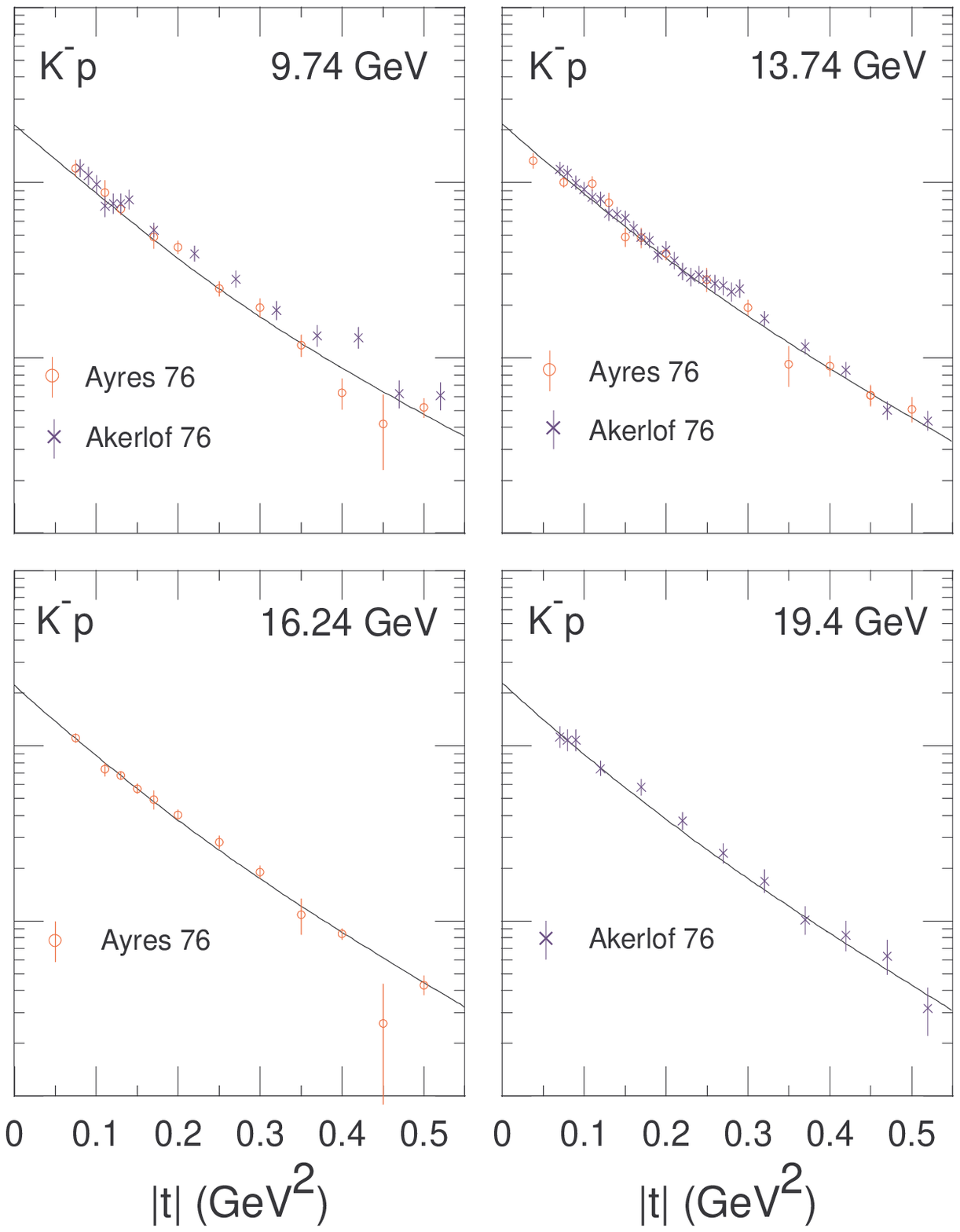}
\caption{$K^{+}p$ and $K^{-}p$ differential cross sections.}
\label{fig:k}
\end{figure}

We also show in Figs. \ref{fig:pp}, \ref{fig:pbarp}, \ref{fig:pi}
and \ref{fig:k} some of the fits to the data. 
We see in Fig.~\ref{fig:pbarp} that our description 
extends very well to S$p\bar p$S energies.
Also, the top-left of Fig.~\ref{fig:pbarp} shows the kind of disagreement that we had to remove: 
the points of Akerlof are in definite
disagreement with those of Ayres. Similar graphs can be plotted for all the data that we removed.
Furthermore, one can see e.g. in the data of Brick \cite{BRICK} in Fig.~\ref{fig:k}
that the first few points are
in strong disagreement with other sets. Such problems explain the rather high value of $|t|_{min}$
that we had to use.

Finally, let us mention that we also considered a fit where one allows the
data of one given set at one given energy to be shifted by a common factor
within one systematic error while treating the statistical error through
the usual $\chi^2$ minimisation. Such a procedure leads to a higher $\chi^2$/d.o.f.,
of the order of 1.15 [7], without affecting the parameters significantly. As
the datasets do not have compatible slopes within the statistical errors, we
preferred to present here the results based on errors added quadratically.
\begin{table}[H]
\begin{center}
\begin{tabular}{|c|c||c||c|}
\hline Quantity     &  Number   & $\chi^{2}/N$& $\chi^{2}/N$  \\
             &  of points  & (soft) & (soft+hard)   \\
\hline \hline
$d\sigma^{pp}/dt$       & 795 &  0.90 & 0.86  \\
$d\sigma^{\bar pp}/dt$  & 226 &  1.01  & 0.99  \\
$d\sigma^{\pi^{+}p}/dt$ & 281 &  0.90  & 0.89  \\
$d\sigma^{\pi^{-}p}/dt$ & 478 &  1.18  & 1.18  \\
$d\sigma^{K^{+}p}/dt$   & 166 &  1.02 & 1.11  \\
$d\sigma^{K^{-}p}/dt$   & 169 &  1.18  & 1.12  \\
\hline Total          &2115 & 1.022 &  0.997 \\
 \hline
\end{tabular}
\end{center}
  \caption{Partial values of $\chi^{2}$, differential cross sections.}\label{tab:chis}
\end{table}

\section{Hard pomeron}
\begin{figure}[H]
\centerline{\includegraphics[scale=1.0]{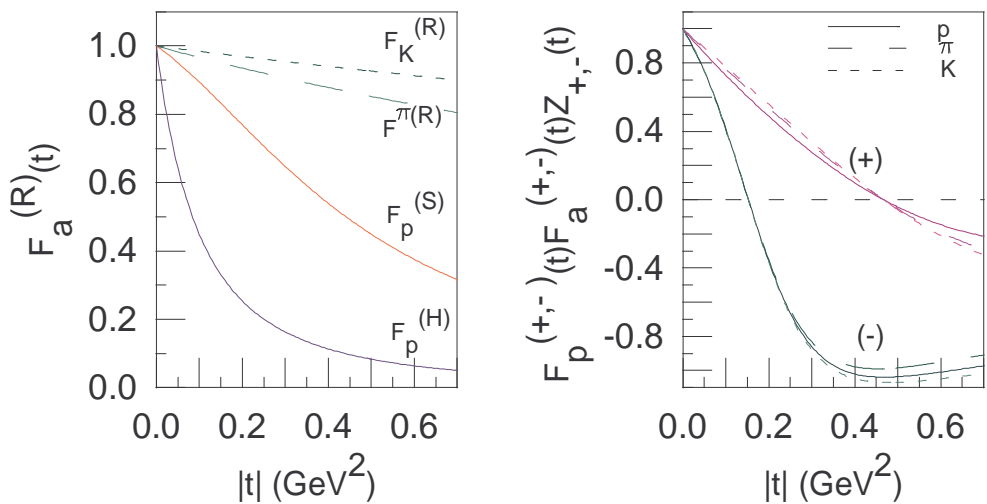}}
\caption{Form factors as function of $|t|$, in the model that includes a 
hard pomeron.} \label{fig:forms}
\end{figure}

One of the motivations of this paper was to confirm the presence of
a small hard component in soft cross sections. The problem however
is that the fit with only one soft pomeron is so good that a hard
component is really not needed here. Following the philosophy of the previous section,
we can nevertheless investigate the effect of its contribution in elastic data
by fixing the parameters from the $t=0$ fit of Table \ref{tab:teq}
and constrain the form factors and trajectories. As can be seen from
Table \ref{tab:chis}, the introduction of a hard pomeron makes the fit
slightly better (the CL rises to about 48\%)
if we allow a different form factor
from that of the soft pomeron in the $pp$ and $\bar p p$ cases.
We obtain the parameters
of the third column of Table~\ref{tab:fits}. The hard pomeron slope
is confirmed to be of the order of 0.1 GeV$^{-1}$, although the errors are large.
We show in Fig. \ref{fig:forms} the form factors of the various trajectories in this case. Note in the $pp$ and $\bar p p$ cases 
that the hard contribution is suppressed at higher $t$ by the form factor.
Forcing it to be identical to the form factor of the soft pomeron
results in a trajectory with a very large slope 
$\alpha'_H\approx 1 GeV^{-2}$. 
\section{Conclusion}
This paper has presented a few advances in the study of elastic cross sections:
\begin{itemize}
\item We have elaborated a complete dataset, including an evaluation 
of the systematic errors for all data. We have shown that statistical
 and systematic errors should be added in quadrature (i.e. the slopes 
of the data from different subsets 
are not consistent if one uses only statistical errors).
\item We have shown that rescattering effects can be neglected in the region 
0.1 GeV$^2\leq |t|\leq 0.5$ GeV$^2$, 6 GeV$\leq \sqrt{s}\leq $ 63 GeV. 
This of course does not necessarily 
mean that the pomeron cuts are small, 
but rather that they can be re-absorbed in a simple-pole parametrisation
\cite{DLel}.

\item We showed that different trajectories must have different form factors. 
We confirm that the crossing-odd meson exchange has a zero. 
We also found evidence for a sharp suppression of the crossing-even 
form factor around $|t|=0.5$ GeV$^2$.
\item The soft pomeron has a remarkably linear trajectory, and leads to a
very good fit that extends well to S$p\bar p$S energies. 
\item Because of the quality of the soft pomeron fit, the elastic data do not
confirm strongly the need for a hard pomeron. It is remarkable however that
the hard pomeron fit gives 0.1 GeV$^{-2}$
for the central value of the slope, in agreement with \cite{DppL}.
\end{itemize}
It is our hope that this dataset, and this study, will serve as a starting
point for precise studies of the whole range of elastic scattering, and especially
for studies of unitarisation effects at higher $s$ or higher $t$, and for the
comparison of several models.
\section*{Acknowledgements}
E.M. acknowledges the support of FNRS (Belgium) for visits to
the university of Li\`ege where part of this work was done.
We thank O.V. Selyugin, L. Szymanowski, M. Polyakov, P.V. Landshoff
and B. Nicolescu for discussion, G. Soyez for partially checking our results,
and A. Prokhudin for help with the data.
\section*{Appendix: experimental data}
\centerline{\large $pp\rightarrow pp$}
\small
\begin{longtable}{|c|c|c|c|c|c|c|c|}
\hline
set&ref.&$\sqrt{s}$&$|t|_{min}$&$|t|_{max}$&syst.&number \\
          &    & (GeV)    & (GeV$^2$) & (GeV$^2$) &&     of points\\
\hline\hline
1001&\cite{AKERLOF}&    9.8  13.8 19.4& 0.075 &  1.03   2.8 3.3&      7\%&50  61 55\\\hline
1002&\cite{ALBROW}&      23.4 26.9 30.6 & 0.15   0.15 0.25 &     1.1 0.55 0.95&     15\%&  19 8 15\\
    &&      32.4 35.2 38.3&    0.20 0.20 0.20&     0.35 0.75 0.7&     &   4 9 9\\\hline
1014&\cite{ALLABY}&       4.5 4.9 5.3&    0.14 0.10 0.27&     2.1 2.7 3.5&     15\%&  24 25 22\\
1015&&       6.2 6.4&    0.058 0.070&     6.0 1.9&      8\%&  37 17\\
1037&&       4.6 4.8 5.0&    2.0 2.2 2.5&     8.6 9.6 10.5&      7\%&  18 15  15\\
&&       5.3 5.8 6.2&    7.6 9.1 9.7&    13 15 17&      &   4  9 4\\
&&       6.5&   11 &    18&      &    4\\
1039&&       6.8&    0.083&     6.7&     10\%&  35\\\hline

1020&\cite{AMALDI}&      23.5 30.7&    0.042 0.016&     0.24 0.11&      1.2\%&  50 48\\
1021&&      30.7 44.7&    0.11 0.05&     0.46 0.29&      2\%&  58 95\\
1030&&      23.5&    0.25&     0.79&      3\%&  28\\
1022&&      23.5 30.7 &    0.83  0.90&     3.0 5.8&      5\%&  34 55\\
&&      44.7  62.5&    0.62 0.27&     7.3 6.3&      &  65 74\\
1023&&      23.5&    3.1&     5.8&     10\%&  21\\
1024&&      30.7&    0.0011&     0.008&      0.40\%&   9\\
1025&&      62.5&    0.0017&     0.009&      0.25\%&  16\\
1026&&      30.7&    0.46&     0.86&      3.5\%&  11\\
1027&&      44.7&    0.001&     0.009&      0.2\%&  24\\
1028&&  44.7 62.5&    0.0092 0.0095&     0.052 0.099&      1\%&  46 49\\\hline
1003&\cite{AMBROSIO}&      52.8&    0.011&     0.048&      {\it 0.4\%\footnote{\label{lum} From the luminosity measurement by the experiment.}}  &  36\\\hline
1009&\cite{AMOS}&      23.5 30.6&    0.0004 0.0005&     0.010 0.018&      1\%&  31 32\\
&&      52.8 62.3&    0.0011 0.0054&     0.055 0.051&      &  34 22\\\hline
1004&\cite{APOKIN}&       9.0 10.0&    0.0019&     0.043 0.05&      1.1\%&  20 18\\\hline
1038&\cite{ARMITAGE}&      53.0&    0.13&     0.46&      5\%&  12\\\hline
1052&\cite{ASAD}&  9.8&    0.825&     3.8&     15\%&  17\\\hline
1005&\cite{AYRES}&       9.8 11.5 13.8&    0.038&     0.75 0.70 0.75&      3\%&  16 17 18\\
&&      16.3 18.2&    0.0375 0.075&     0.80 0.75&      &  19 15\\\hline
1006&\cite{BEZNOGIKH}& 4.4 5.1 5.6&0.0008 0.0092 0.0089& 0.013 0.10 0.11 & {\it 2\%\footnote{\label{opt} From the uncertainty on the optical point used to normalise the data.}} &  34 22 27\\
&&       6.1 6.2 6.5&    0.0009 0.0011 0.015&     0.11 0.014 0.11&          &  67 35 30\\
&&       6.9 7.3 9.8&    0.011 0.0093 0.0010&     0.11 0.11 0.12 &          &  26 33 66\\
&&       7.7 8.0 8.3&    0.011 0.0171 0.0093&     0.11 0.11 0.11&          &  29 24 28\\
&&       8.6 8.7 8.8&    0.0009 0.0011 0.0009&     0.11 0.015 0.11&         &  65 47 65\\
&&       9.3 10.0 10.2&    0.0114 0.0109 0.0108&     0.12&         &  29 34 29\\
&&      10.3 10.4 10.6&    0.0008 0.013 .0008&0.015 0.12 0.015& &  37 35 44\\
&&10.7 11.0 11.2&0.0108 0.013 0.011& 0.12 0.12 0.12 & &  33 33 30\\
&&      11.5&    0.011 0.0010&     0.12 0.11&       &  26 156\\\hline
1013&\cite{BRANDENBURG}&       4.6&    0.023&     1.5&      2\%&  97\\\hline
1031&\cite{BREAKSTONE}&      31.0 53.0 62.0&    0.050 0.11 0.13&     0.85&   10\% & 24 24 23\\
1064&&      53.0&    0.62&     3.4&     20\%&  31\\\hline
1055&\cite{BRICK}&      16.7&    0.01&     0.62&        {\it 2\%\footnote{\label{brick} This uncertainty in the luminosity, originally included in the statistical error, has been removed from it.}}  &  26\\\hline
1007&\cite{BURQ}&      13.8 16.8 &    0.0022&     0.039&      1\%&  73 68 \\
&&      21.7 23.8&    &     &      &  64 60\\\hline
1054&\cite{COOL}&      13.8 19.4&    0.035&     0.095&      0.8\%&   7 7\\\hline
1058&\cite{CONETTI}&      19.5 27.4&    5.0 2.3&    12 16&     20\%&  31 87\\\hline
1017&\cite{DALKHAZHAV}&       4.7&    0.0028&     0.14&  {\it 1.6\%}$^{\ref{opt}}$    &  13\\\hline
1053&\cite{DEVENSKI}&       9.8&    0.012&     0.12&     {\it 3\%}$^{\ref{opt}}$  &  10\\\hline
1042&\cite{DIDDENS}&       5.0&    0.011&     0.34&     15\%&   5\\
1044&&       5.6&    0.019&     0.56&     13\%&   5\\
1045&&       6.1 7.1&    0.036 0.064&     0.79 1.0&     20\%&   5 4\\
1046&&       6.5&    0.032&     1.1&     17\%&   5\\\hline
1019&\cite{EDELSTEIN}&       4.5 5.5&    0.016 0.027&     5.1 4.9&     15\%&  31 32\\
&&       6.3 7.6&    0.032 0.079&     3.8 2.8&     &  30 29 \\\hline
1029&\cite{ERHAN}&      53.0&    0.64&     2.05&     10\%&  15\\\hline
1057&\cite{FAISSLER}&      19.5 27.4&    5.0 5.5&    12 14&     15\%&  34 30\\\hline
1056&\cite{FIDECARO}&      19.4&    0.61&     3.9&        {\it 15\%}\footnote{This uncertainty is the same as in \cite{RUSACK}.}  &  33\\\hline
1016&\cite{FOLEY}&       4.7 5.1 5.4&    0.058 0.049 0.066&     0.82 0.86 0.78&      5\%&  13 13 12\\
&&       5.8 6.2&    0.042 0.12 &     0.70 0.81&    &  12 11\\
1018&&   4.7 5.5 6.2 & 0.2 0.22 0.23&     0.89 0.74 0.79&      5\%&   9 7 7\\
&&       6.5 6.9&    0.24 0.25&     0.81 0.75&      &   7 6\\\hline
1048&\cite{GESHKOV}&       7.6 9.8 11.5&    0.0027 0.0026 0.0028&     0.119 0.12 0.12&    {\it 2\%}$^{\ref{opt}}$      &  21 23 21\\\hline
1049&\cite{RUSACK}&       8.2 10.2 11.1&    0.29 0.34 0.34&     1.93 1.98 1.98&     15\%&  21 20 20\\
&&      12.3 13.8 15.7&    0.35&     0.70 2.0 0.99&    &   8 19 11\\
&&      16.8 17.9 18.9&    0.35 0.35 0.29&     2.1&   &  32 29 30\\
&&      19.9 20.8 21.7&    0.29&     2.1 2.0 2.0 &    &  29 19 17\\\hline
1043&\cite{HARTING}&       5.0 6.0&    0.13 0.19&     2.0 3.6&      7\%&  22 20\\\hline
1040&\cite{JENNI}&       4.5&    0.0018&     0.097&      1\%&  55\\\hline
1050&\cite{BRUNETON}&       9.2&    0.16&     2.0&      {\it 2\%}$^{\ref{opt}}$&  27\\\hline
1036&\cite{KUZNETSOV}&      10.0&    0.0006&     0.031&      0.9\%&  72\\
1035&&      12.3&    0.0007&     0.029&      0.69\%&  58\\
1034&&      19.4&    0.0007&     0.032&      0.56\%&  69\\
1033&&      22.2&    0.0005&     0.030&      0.57\%&  63\\
1032&&      23.9&    0.0007&     0.032&      0.5\%&  66\\
1008&&      27.4&    0.0005&     0.026&      0.52\%&  60\\\hline
1010&\cite{NAGY}&      52.8&    0.83&     9.8&      5\%&  63\\\hline
1041&\cite{OREAR}&       4.9&    1.2&     2.5&     10\%&   5\\\hline
1011&\cite{RUBINSTEIN}&      13.8 19.4&    0.55 0.95&     2.5 10.3&     15\%&  20 35\\\hline
1012&\cite{SCHIZ}&      19.4&    0.021 &     0.66&     {\it 4\%}\footnote{\label{sch}The $t$-dependent systematics have been included in the statistical error.}&  134   \\
\hline
\end{longtable}
\centerline{\large$\bar pp\rightarrow \bar pp$}
\begin{longtable}{|c|c|c|c|c|c|c|c|}
\hline
set&ref.&$\sqrt{s}$&$|t|_{min}$&$|t|_{max}$&syst.&number \\
          &    & (GeV)    & (GeV$^2$) & (GeV$^2$) &&     of points\\\hline\hline
1130&\cite{ABE}& 546.0 &0.026 & 0.078 &   {\it 0.52\%}\footnote{\label{abe}From Table VI of \cite{ABE}.}       &  14 \\
1132&& 1800.0&0.035& 0.285&   {\it 0.48\%}$^{\ref{abe}}$  &  26\\\hline
1101&\cite{AKERLOF}&   9.8 13.8 19.4&    0.075&     1.0 0.95 0.75&      7\%&  31 30 13\\\hline
1102&\cite{AMBROSIO}&      52.8&    0.011&     0.048&        {\it 1.54 \%}$^{\ref{lum}}$  &  48\\\hline
1103&\cite{AMOS}&  30.4 52.6 &    0.0007 0.001 &     0.016 0.039 &      2.5\%&  29 28 \\
&&  62.3 &   0.0063 &     0.038 &      &   17 \\
1104&&  1800.0&   0.034&     0.63&  9\%     &   17 51\\\hline
1105&\cite{ANTIPOV}&6.9 7.0 8.8&0.19 0.83 0.075&     0.58 3.8 0.58&5\%&  22 17 33\\\hline
1106&\cite{ARNISON}&     540.0&    0.045&     0.43&      8\%&  36\\\hline
1107&\cite{ASAD}&       7.6 9.8&    0.53  0.83 &     5.4  3.8&     15\%&  30  17\\
\hline
1108&\cite{AYRES}&       9.8 11.5 13.8&    0.038&     0.75 0.5 0.75&      3\%&  17 13 15\\
&&      16.3 18.2&    0.075 0.038&     0.6&      & 11 13\\\hline
1109&\cite{BATYUNYA}&       6.6&    0.055 &     0.88&        {\it 2.1 \%}$^{\ref{opt}}$  &  43\\\hline
1110&\cite{BERGLUND}&       4.6&    0.19  &     3.0&      5\%&  35\\\hline
1111&\cite{BERNARD}&     546.0&    0.0022&     0.035&      2.5\%&  66\\
1112&&     630.0&    0.73  &     2.1&     15\%&  19\\\hline
1126&\cite{BIRNBAUM}&       5.6&    0.11 &     1.3&        {\it 10\%}\footnote{From \cite{BIRNBAUM}.}  &  23\\\hline
1114&\cite{BOGOLYUBSKY}&       7.9&    0.055 &     1.0&        {\it 0.8\%}$^{\ref{opt}}$   &  52\\\hline
1113&\cite{BOZZO}&     546.0&    0.032&     0.50&      5\%&  87\\
1117&&     546.0&    0.46  &     1.5&     10\%&  34\\\hline
1118&\cite{BRANDENBURG}&       4.6&    0.023&     1.5&      2\%&  97\\\hline
1115&\cite{BREAKSTONE}&      53.0&    0.52 &     3.5&     30\%    &  27\\
1116&&      31.0 53.0 62.0&    0.05 0.11 0.13 & 0.85 & 15\%  &  22 24 23\\\hline
1128&\cite{COOL}&      13.8 19.4&    0.035 &     0.095&      0.8\%&   7 7\\\hline
1129&\cite{ERHAN}&      53.0&    0.64  &     1.9&     10\%&   8\\\hline
1124&\cite{FOLEY}&       4.5 4.9&    0.03 0.043 &     0.18 0.52&      5\%&   6 10\\
1125&&       4.9 5.6&    0.20 0.22&     0.49 0.45&      5\%&   5 4\\\hline
1123&\cite{JENNI}&       4.5&    0.0018&     0.097&      1\%&  55\\\hline
1127&\cite{BRUNETON}&       8.7&    0.17 &     1.24&      {\it 2\%}$^{\ref{opt}}$ &  11\\\hline
1119&\cite{LEWIN}&       7.9&    0.07  &     0.62&      {\it 2\%}$^{\ref{opt}}$    &  23\\\hline
1131&\cite{OWEN}&       4.5&    0.76&     5.5&      5\%&  10\\\hline
1121&\cite{RUSS}&       5.6&    0.085 &     1.2&      5\%&  34\\\hline
1120&\cite{RUBINSTEIN}&      13.8&    0.55  &     2.5&     15\%&  15 \\
1122&&      19.4&    0.95  &     3.8&     35\%&   7\\\hline
\end{longtable}

\centerline{\large $\pi^+ p\rightarrow \pi^+ p$}
\begin{longtable}{|c|c|c|c|c|c|c|c|}
\hline
set&ref.&$\sqrt{s}$&$|t|_{min}$&$|t|_{max}$&syst.&number \\
   &    & (GeV)    & (GeV$^2$) & (GeV$^2$) &&     of points\\\hline\hline
1212&\cite{ADAMUS}& 21.7&  0.08  &   0.94&  {\it 2\%}$^{\ref{opt}}$   & 18\\\hline
1205&\cite{AKERLOF}&9.7 13.7 19.4&    0.075 &     1.7 1.7 1.8&      7\%&  70 63 53\\\hline
1203&\cite{APOKIN}&       9.0 9.9&    0.002 0.0019&     0.043 0.05&      1.1\%&  20 18\\\hline
1214&\cite{AZHINENKO}&       7.8&    0.075 &     0.68&    {\it 1.4\%}$^{\ref{opt}}$       &  13\\\hline
1206&\cite{ASAD}&       9.7&    0.75  &     3.9&     15\%&  22\\\hline
1207&\cite{AYRES}&       9.7 11.5&    0.038&     0.8  0.7&      3\%&  19 17\\
&&      13.7 16.2 18.1&    0.11 0.038 0.075  &     0.8&      &  17 19 18\\\hline
1215&\cite{BAGLIN}&       4.4&    0.46  &    17.3&     15\%&  84\\\hline
1201&\cite{BRANDENBURG}&       4.5&    0.023&     1.5&      2\%&  97\\\hline
1210&\cite{BRICK}&      16.6&    0.01  &     0.58&        {\it 2\%}$^{\ref{brick}}$  &  25\\\hline
1209&\cite{COOL}&      13.7  19.4&    0.035 &     0.095&      0.8\%&   7 7\\\hline
1204&\cite{BRUNETON}&       9.2&    0.16 &     1.92&    {\it  2\%}$^{\ref{opt}}$ &  18\\\hline
1202&\cite{RUBINSTEIN}&       5.2&    0.65  &     3.8&     10\%&  24\\
1208&&      13.7 19.4&    0.55 0.95  &     2.5 3.4&     15\%&  20 20\\\hline
1211&\cite{SCHIZ}&      19.4&    0.022 &     0.66&      {\it 4\%}$^{\ref{sch}}$&  133\\
\hline
\end{longtable}
\centerline{\large $\pi^- p\rightarrow \pi^- p$}
\begin{longtable}{|c|c|c|c|c|c|c|c|}
\hline
set&ref.&$\sqrt{s}$&$|t|_{min}$&$|t|_{max}$&syst.&number \\ 
   &    & (GeV)    & (GeV$^2$) & (GeV$^2$) &&     of points\\\hline\hline
1302&\cite{AKERLOF}&      9.7 13.7 19.4 &    0.075 &    1.60 1.83 2.38&7\%&64  60 61\\\hline
1310&&       6.9 8.7&    0.075 &     0.78 0.70&      5\%&  38 38\\
1324&&       8.7&    0.19  &     1.3&     10\%&  28\\\hline

1301&\cite{APOKIN}&       8.7&    0.002 &     0.008&      1.5\%&  21\\
1312&&       8.0 8.4 8.7&    0.0012 0.0015 0.0016&     0.025 0.03 0.034&      1.5\%&  19 19 36 \\
&&       9.3 9.8&    0.0022 0.0028&     0.05 0.056&      &  17 18\\
&&      10.4 10.6&    0.0035 0.0014&     0.077 0.085&    &  18 19\\
1314&&       8.7 9.7&    0.0016 0.0022&     0.021 0.035&      {\it 1\%}$^{\ref{opt}}$   &  20 34\\\hline
1309&\cite{ASAD}&       6.2 9.7&    0.65 0.73 &     6.0 7.8&     15\%&  22 46\\\hline
1315&\cite{AYRES}&       9.7 11.5&    0.038&     0.75 0.50&      3\%&  18 13\\
&&      13.7 16.2 18.1&    0.038&     0.80 0.75 0.80&  &  19 18 19\\\hline
1304&\cite{BAGLIN}&       6.2 7.6&    7.4 10.  &    17 25&     15\%&   6  4\\\hline
1305&\cite{BRANDENBURG}&       4.5&    0.023&     1.5&      2\%&  97\\\hline
1318&\cite{BURQ}&      13.7 16.8 19.4&    0.0022 0.0022 0.0023 &     0.039&      1\%&  73 68 64\\
&&      21.7 23.7 24.7&    0.0022&     &      & 116 59 56\\
&&      25.5&    &     0.038&      &  57\\\hline
1317&\cite{CHAPIN}&      13.7&    0.028 &     0.092&      {\it 10\%}\footnote{From \cite{CHAPIN}.}  &   5\\\hline
1303&\cite{COOL}&      13.7 19.4&    0.035 &     0.095&      0.8\%&   7 7\\\hline
1308&\cite{CORNILLON}&       5.2&    0.75  &     4.5&      9\%&  25\\
1325&&       6.6&    0.3   &     5.2&     12\%&  44\\\hline
1311&\cite{DEREVSHCHIKO}& 7.9 8.2 8.9& 0.057 0.16 0.066& 0.20 0.49 0.37&5\%&  14 18 25\\
&& 9.3 9.6 9.8& 0.068 0.04 0.082&  0.42 0.37 0.55&&  18 25 27\\
&&      10.2 10.2&    0.054 0.055 &     0.53 0.46&    &  19 17\\
1306&&       9.7&    0.035 &     0.40&      2.5\%&  37\\\hline
1326&\cite{DZIERBA}&       5.2&    0.015 &     0.77&      6\%&  41\\\hline
1307&\cite{HARTING}&       4.1 4.9 6.0&    0.05  0.09 0.19 &     1.1 2.0  3.6&      7\%&  23 24 20\\\hline
1320&\cite{JENKINS}&       4.02 4.06 4.11&    4.5   &     9.3 9.9 9.9&      3\%&  25 28 28\\
&&       4.14 4.18 4.21&    4.9   &     9.9 10.1 10.9&      &  26 27 30\\
&&       4.26 4.30 4.33&    5.3   &    10.7  10.5 10.7&     &  26 22 21\\\hline
1313&\cite{BRUNETON}&       8.6&    0.17 &     2.1&      {\it 2\%}$^{\ref{opt}}$ &  20\\\hline
1321&\cite{OREAR}&       4.8&    1.2   &     2.4&     10\%&   4\\\hline
1322&\cite{RUSS}&       5.6&    0.15 &     1.8&      5\%&  38\\\hline
1316&\cite{RUBINSTEIN}&      13.7 19.4&    0.55  0.95 &     2.5 10&     15\%&  20 31\\\hline
1319&\cite{SCHIZ}&      19.4&    0.021&     0.66&      4\%& 134\\\hline
\end{longtable}
\newpage
\centerline{\large $K^- p\rightarrow K^- p$}
\begin{longtable}{|c|c|c|c|c|c|c|c|}
\hline
set&ref.&$\sqrt{s}$&$|t|_{min}$&$|t|_{max}$&syst.&number \\
   &    & (GeV)    & (GeV$^2$) & (GeV$^2$) &&     of points\\\hline\hline
1414&\cite{ADAMUS}&      21.7&    0.12  &     0.94&    {\it 2\%}$^{\ref{opt}}$  &  17\\\hline
1406&\cite{AKERLOF}&       9.7 13.7 19.4&    0.075 0.075 0.07&     1.5 1.9 1.9&      7\%&  21 35 35\\\hline
1404&\cite{APOKIN}&       9.0 10.0&    0.0019&     0.043 0.050&      1.1\%&  20 18\\\hline
1408&\cite{ASAD}&       9.7&    0.75  &     7.0&     15\%&  23\\\hline
1407&\cite{AYRES}&       9.7 11.5&    0.038 &     0.70 0.65 &      3\%&  16 16\\
&&      13.7 16.2 18.2&    0.075 0.075 0.038&   0.75 0.70 0.75&    &  13 16 17\\\hline
1415&\cite{BARTH}&      11.5&    0.090  &     0.98&        {\it 2.6\%}\footnote{From the error on the topological cross section used to normalise the data.}  &  36\\\hline
1411&\cite{BRICK}&      16.6&    0.02  &     0.56&        {\it 2\%}$^{\ref{brick}}$  &  10\\\hline
1402&\cite{BRANDENBURG}&       4.5 5.2&    0.023&     1.5&      2\%&  97 97\\\hline
1409&\cite{COOL}&      13.7&    0.045 &     0.095&      {\it 0.8\%}$^{\ref{opt}}$ &   6\\\hline
1405&\cite{BRUNETON}&       9.2&    0.16 &     1.25&      2\%$^{\ref{opt}}$ &  13\\\hline
1401&\cite{LEWIN}&       7.8&    0.09  &     1.4&     {\it 2\%}$^{\ref{opt}}$      &  48\\\hline
1410&\cite{RUBINSTEIN}&      13.7 19.4&    0.55  0.95&     2.1  2.4&     15\%&  16 12\\
1403&&       5.2&    0.75  &     2.2&     10\%&  12\\\hline
\end{longtable}
\centerline{\large $K^- p\rightarrow K^- p$}
\begin{longtable}{|c|c|c|c|c|c|c|c|}
\hline
set&ref.&$\sqrt{s}$&$|t|_{min}$&$|t|_{max}$&syst.&number \\
   &    & (GeV)    & (GeV$^2$) & (GeV$^2$) &&     of points\\\hline\hline
1508&\cite{ANTIPOV}&       7.0 8.7&    0.075 &     0.78&      5\%&  38 38\\
1513&&       8.7&    0.19 &     1.3&      10\%&   28\\
\hline
1507&\cite{ASAD}&       6.2&    0.65  &     4.25&     15\%&  16\\\hline
1511&\cite{AYRES}&       9.7 11.5 13.7&    0.075 0.0375 0.0375&     0.75 0.45 0.75&      3\%&  14 12 16\\
&&      16.2 18.2&    0.075 &     0.6 0.75&      &  13 15\\\hline
1510&\cite{AKERLOF}&       9.7 13.7 19.4&    0.070  &     1.4 1.7 1.0&7\%&  26 42 17\\\hline
1501&\cite{BERGLUND}&       4.5&    0.19  &     2.3&      5\%&  49\\\hline
1503&\cite{BRANDENBURG}&       4.5 5.2&    0.023 &     1.5&      2\%&  97 97\\\hline
1502&\cite{CAMPBELL}&       4.5&    0.0070 &     2.1&        {\it 1.8\%}$^{\ref{opt}}$   &  42\\\hline
1505&\cite{DEBOER}&       5.3&    0.010  &     2.4&     {\it 2\%}$^{\ref{opt}}$      &  27\\\hline
1506&\cite{DREVILLON}&       5.3&    0.045 &     1.9&   {\it 2\%}$^{\ref{opt}}$    &  62\\\hline
1509&\cite{BRUNETON}&       8.6&    0.17 &     2.0&      {\it 2\%}$^{\ref{opt}}$ &  13\\\hline
1504&\cite{MILLER}&       5.3&    0.035 &     1.3&      3\%&  41\\\hline
1512&\cite{RUBINSTEIN}&      13.7 19.4&    0.55 0.95 &     2.5 2.2&     15\%&  20  8\\\hline
\end{longtable}

\end{document}